\documentclass[12pt]{iopart}
\usepackage[utf8]{inputenc}
\usepackage[square,numbers]{natbib}
\usepackage{graphicx}
\usepackage{color}
\usepackage{bm}
\usepackage{amssymb}
\usepackage[normalem]{ulem}
\usepackage[dvipsnames]{xcolor}
\usepackage[colorlinks,urlcolor=NavyBlue,citecolor=NavyBlue,linkcolor=NavyBlue,pdfusetitle]{hyperref}

\usepackage{mathrsfs,wasysym,amssymb}
\usepackage{textcomp, gensymb}
\usepackage{graphicx}
\usepackage{ulem}
\usepackage{epstopdf}
\usepackage{color}
\usepackage{hyperref}

\begin{document}


\title{Understanding supernova gravitational waves with protoneutron star asteroseismology}

\author{Hajime Sotani$^{1,2,3}$}

\address{$^1$Department of Mathematics and Physics, Kochi University, Kochi, 780-8520, Japan}

\address{$^2$RIKEN Center for Interdisciplinary Theoretical and Mathematical Sciences (iTHEMS), RIKEN, Wako 351-0198, Japan}

\address{$^3$Theoretical Astrophysics, IAAT, University of T\"{u}bingen, 72076 T\"{u}bingen, Germany.}



\begin{abstract}
Supernovae are one of the most promising gravitational wave sources. But, since the system of the supernovae is nearly spherically symmetric, the expected gravitational waves from them are relatively weak, compared to the case of the compact binary mergers. Thus, at least using the current gravitational wave detectors, only the gravitational waves from a supernova that occurred in our galaxy could be detected. To reliably extract information from gravitational waves originating from such a low event rate, thorough preparation is essential. However, because supernova gravitational waves strongly depend on model parameters, such as progenitor mass and the equation of state for dense matter, it may be difficult to extract physical properties even if the gravitational waves are detected. The universal relations between gravitational-wave signals and physical properties, independent of model parameters, are important for solving this difficulty. To discuss such a universal relation, in this article, we systematically examine the protoneutron-star oscillation frequencies with the linear analysis, the so-called asteroseismology, and compare them with the gravitational wave signals in the simulations.
\end{abstract}

\section{Introduction}
\label{sec:introduction}

Gravitational waves have become a new means of observing celestial bodies, referred to as multi-messenger astronomy, together with electromagnetic waves and neutrinos. In practice, the gravitational waves detected at the GW170817 event enable us to constrain the mass, radius, and tidal deformability of neutron stars before merger~\cite{GW170817,GW170817_multi}. During the fourth observing run by the LIGO-Virgo-KAGRA collaborations, which concluded in November 2025, more than 300 gravitational wave events have been detected. Now, the detectors are updating, and an additional six-month observing run is also planned to begin in the late summer or early fall of 2026, utilizing the available detectors. In the future, as the accuracy of detectors improves further, gravitational waves from sources other than compact binary mergers may be detected. In particular, supernova explosions, which happen at the end of massive stars' lives, are one of the promising sources~\cite{Ott09,Abdikamalov22,SZ22,Arimoto23,MZ24,Muller26}. However, the gravitational waves from supernova explosions are much weaker than those from compact binary mergers, because the system of supernova explosions is roughly spherically symmetric, unlike compact binary mergers. Thus, one has to thoroughly prepare theoretical studies before gravitational waves from a supernova explosion are really detected. 

Theoretical studies of gravitational waves from core-collapse supernovae have primarily been conducted through numerical simulations. These attempts showed the emergence of a “ramp-up” signal after core bounce as a primary gravitational-wave signal, whose frequencies increase from a few hundred Hz up to the kHz range within around 1 second, e.g.,~\cite{Muller13,Pablo13,Kuroda16,MRBV2018,Jakobus23,Mezzacappa23,Vartanyan23}. This signal is originally considered as a result of the Brunt-V\"{a}is\"{a}l\"{a} frequency at the protoneutron star surface (or surface gravity ($g$-) mode)~\cite{Muller13,Pablo13}. Subsequently, through linear analysis, it is found that this ramp-up signal comes from the fundamental ($f$-) mode oscillations (or the $g$-mode oscillations differing from the standard classification) excited in the protoneutron stars produced via core-collapse supernovae, e.g.,~\cite{MRBV2018,TCPF18,TCPOF19,TCPOF19b,SKTK2019,ST2020a,STT2021,Bizouard21,Wolfe23,Bruel23,SMT24,SMT25}. In addition to the ramp-up mode, a gravitational-wave signal at approximately 100 Hz, which is almost constant in time, induced by the steady accretion shock instability, e.g.,~\cite{Kuroda16,Murphy09,Andresen17,Andresen19,Vartanyan20} and a signal resembling the $g_1$-mode excited in the protoneutron star with a frequency decreasing over time~\cite{MRBV2018,Jakobus23} has also been reported, although these signals appear to be highly dependent on the supernova models.
In this article, we also focus only on the gravitational wave signals produced after $\sim$ 1 second after core bounce. 

To understand each component of the gravitational-wave signal that appears after core bounce in the simulations, linear analysis of the protoneutron star is quite important. This technique is known as (gravitational wave) asteroseismology~\cite{KS99,Nils21}, which is similar to seismology on Earth and helioseismology on the Sun. Since an object has its own eigenfrequencies depending on the interior properties, one could extract the properties of the object as an inverse problem by observing such frequencies~\cite{AK96,AK98}. In practice, by identifying the quasi-periodic oscillations observed in magnetars with the crustal torsional oscillations, the neutron star mass, radius, and crust equation of state (EOS) are constrained, e.g.,~\cite{SW09,GNHL2011,SNIO2012,SIO2019,SKS23,Sotani24a}. Similarly, it is suggested that one may extract the neutron star mass, radius, EOS, and rotational properties once one detects the gravitational wave frequencies from a neutron star, e.g.,~\cite{AK96,AK98,SKH2004,SYMT2011,PA2012,DGKK2013,KHA15,Sotani23}. This technique is also useful even for the protoneutron stars produced after core-collapse supernovae, e.g.,~\cite{MRBV2018,SKTK2019,ST2020a,TCPF18,TCPOF19,TCPOF19b,STT2021,SS2019}.

Once supernova gravitational waves are detected, they help for revealing the EOS of dense matter. In fact, the central density can exceed roughly twice the nuclear saturation density even in canonical-mass neutron stars, while it may reach values larger by a factor of five or more in massive $2M_\odot$ neutron stars. In addition, although comparable densities may be achieved even on Earth transiently in heavy-ion collisions, reproducing neutron-star conditions on Earth is quite difficult due to the extreme isospin asymmetry, the long-lived nature of the state, and the possible appearance of additional degrees of freedom, such as hyperons and deconfined quarks~\cite{ST83}. With these reasons, 
the constraint on the EOS for a higher density region can be possible only via observations of compact stars, although one might get the constraint on the EOS for a lower density region through the terrestrial experiments~\cite{SNN22,SO22}. In fact, the discoveries of massive neutron stars exclude the soft EOS, with which the expected maximum mass does not reach the highest mass observed so far~\cite{D10,F21,Saffer25,Romani26}. Even so, since the observation of a (cold) neutron star gives us only one set of mass and radius, to constrain the EOS, one may need to observe several different objects. Meanwhile, since the mass and radius of a protoneutron star evolve with time due to mass accretion, one can obtain the sequence of the mass and radius of the protoneutron star, even if one observes only one event of a supernova gravitational wave.

The supernova gravitational waves depend on the model parameters, such as the EOS and progenitor mass. Therefore, the universal relation, if any, is important for extracting information from the gravitational waves, where the universal relation is the relation between the combinations of the protoneutron star properties and gravitational wave frequencies independently of the model parameters. For a cold neutron star, several universal relations have been discovered, e.g.,~\cite{AK96,AK98,TL2005,SK21,Sotani26}. In the case of supernova gravitational waves as well, the universal relations have been derived, where the gravitational wave frequencies are expressed as a function of the protoneutron star average density or surface gravity \cite{TCPF18,STT2021}. 

We note that, for hot neutron stars (sometimes referred to as proto-neutron stars), assuming a constant (redshifted) temperature ($\gtrsim 1$ MeV) or constant entropy per baryon ($\gtrsim 1$ k$_B$), several studies have been conducted on their oscillation frequencies, including those in the context of universal relations, 
e.g., \cite{LP22,Lozano22,Kumar24,Barman25,Deepak25,Zheng25,Guha25}. Also, there are some studies on how the different nuclear properties and composition impact the static properties with the same assumption, e.g.,~\cite{Sen21,BZ24,Tsiopelas24,Wu25,Togashi25,Kunkel25}. In fact, the temperature of a newly born proto-neutron star is of order 10-50 MeV (corresponding to $\sim (1-5)\times 10^{11}$ K), e.g., \cite{BL86,Prakash97,Pons99}. However, protoneutron stars produced via core-collapase supernovae are thermally non-relaxed, exhibiting strong radial temperature gradients, especially between the core and the crust, and the stellar interior becomes nearly isothermal after thermal relaxation within $\sim 10^2-10^3$ years, where the red-shifted temperature becomes up to $\sim 10^8$ K (corresponding to $\sim 10$ keV), and the entropy per baryon becomes around $10^{-3}$ $k_{\rm B}$, e.g., \cite{Gnedin01,Yakovlev04}. But, unfortunately, the properties of the (worm) neutron stars with $\sim 10^8$ K are almost the same as those for cold neutron stars. Thus, it is quite unclear to what extent studies conducted under the assumption that the temperature is constant and $\gtrsim 1$ MeV, or that the entropy per baryon is constant and $\gtrsim 1$ $k_{\rm B}$, reflect realistic conditions.

To discuss the universal relation in supernova gravitational waves, there are still uncertainties in the treatment of gravity in the numerical simulations. Namely, some simulations have been conducted within Newtonian gravity, adopting the effective gravitational potential that mimics the Tolman-Oppenheimer-Volkoff solution (hereafter referred to as effective GR)~\cite{Rampp02,effectiveGR}, while other simulations have also been conducted in the relativistic framework (hereafter referred to as GR). Moreover, independent of the adopted gravitational theory, the simulation results also depend on the treatment of the dimension of gravity. If the simulations are conducted in multi-dimensions, not only the fluid but also the gravitational potential in the Newtonian case or metric in the relativistic case should be consistently calculated in the multi-dimensions. However, for simplicity, the gravitational potential or metric is kept in spherical symmetry during the simulation in some cases. We call these simulations with the assumption that the gravitational potential or metric is kept in spherical symmetry ``monopole gravity" in this manuscript. On the other hand, the simulations where the gravitational potential or metric is consistently treated as the same dimension as in fluid motion are called ``multipole gravity" in this manuscript. In practice, since we consider only two-dimensional simulations in this manuscript, the simulations with multipole gravity are those where the gravitational potential or metric is also considered in the two-dimensional space. We note that we did not consider the rotational (or magnetic) effects in the simulations, but if one takes into account such effects, the treatment of monopole gravity may not be suitable, especially for fast-rotating and/or strongly magnetized cases. So, in this article, we discuss how robust the universal relations expressing the supernova gravitational waves are with respect to the difference in the gravitational theory with monopole gravitational potential, i.e., the effective GR or GR, and with respect to the dimension of gravity in GR simulations, i.e., monopole or non-monopole (two-dimensional) potential.

This article is organized as follows. In Sec.~\ref{sec:PNS}, we describe the protoneutron star models provided using the data obtained by the numerical simulations, which become a background model to make a linear analysis. In Sec.~\ref{sec:LA}, we determine the specific frequencies of the protoneutron stars by solving the eigenvalue problem and see the behavior of frequencies. Then, in Sec.~\ref{sec:UR-eGR}, we discuss the universal relations in the supernova gravitational waves in the effective GR, while, in Sec.~\ref{sec:UR-GR}, we discuss how the specific frequencies corresponding to the gravitational-wave signal appearing in the simulations depend on the gravitational theory and/or the dimension of gravity together with the universal relations expressing the supernova gravitational waves. Finally, we conclude this study in Sec.~\ref{sec:conclusion}. Unless otherwise mentioned, we adopt geometric units with $c=G=1$, where $c$ and $G$ denote the speed of light and the gravitational constant, and use the metric signature $(-,+,+,+)$.

\section{Protoneutron star models}
\label{sec:PNS}

\begin{figure}
 \centering
 \includegraphics[scale=0.5]{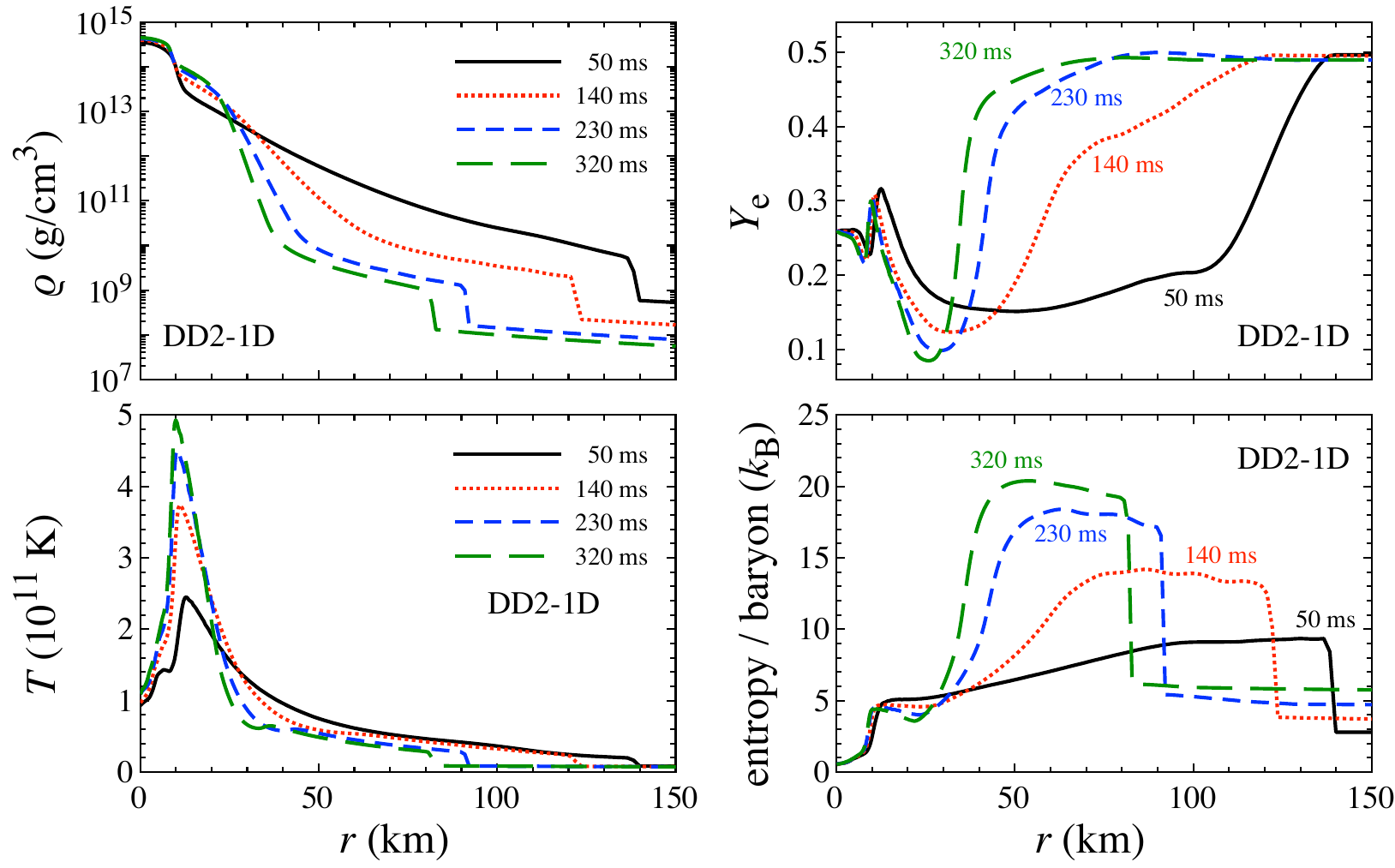}
 \caption{The evolution of radial profile for the density (top-left panel), electron fraction (top-right panel), temperature (bottom-left panel), and entropy per baryon (bottom-right panel). The different lines correspond to the different times after core bounce. The simulation is conducted with DD2 EOS in 1D-effective GR. Modified figure in Ref.~\cite{ST2020b}. }
\label{fig:rho_t_DD2}
\end{figure}

To determine the eigenfrequencies of protoneutron stars through linear analysis, one first has to prepare the background models for protoneutron stars. Unlike (cold) neutron stars, wghere one can construct the stellar models using a zero temperature EOS, i.e., the relation between the energy density and pressure, one needs to know the radial profiles of density ($\rho$), pressure ($p$), temperature ($T$) (or entropy per baryon), and electron fraction ($Y_e$) for constructing protoneutron star models with a finite temperature EOS. However, such profiles are determined after numerical simulations for core-collapse supernovae. This is one reason why asteroseismology of protoneutron stars has not been pursued as actively.  In this study, we provide the protoneutron star models using the data obtained from the numerical simulations in the effective GR or GR.
The simulations in the effective GR have been conducted by neutrino radiation hydrodynamic simulations in \texttt{3DnSNe} code~\cite{Takiwaki16,OConor18,Kotake18,Nakamura19,Sasaki20,Zaizen20}, while the GR simulations have been done with the extended conformal flatness condition (xCFC) \cite{xCFC} for the
space-time metric, using the \texttt{COCONUT-FMT} code~\cite{{DFM02,Muller10,Muller12,Muller15}}.
In particular, we consider the linear analysis on the spherically symmetric background models, assuming that the protoneutron stars are static at each time step. So, if the numerical simulations are done in multidimension, the protoneutron star models are constructed by averaging in the angular direction and by neglecting the radial motion. Then, as shown in Fig.~\ref{fig:rho_t_DD2}, the radial profiles of properties are obtained, depending on the time after core bounce, where the numerical simulation is conducted adopting DD2 EOS~\cite{DD2} with the effective GR~\cite{ST2020b}. We note that all the simulations with the effective GR discussed in this article are conducted with the monopole gravitational potential.

\begin{figure}
 \centering
 \includegraphics[scale=0.5]{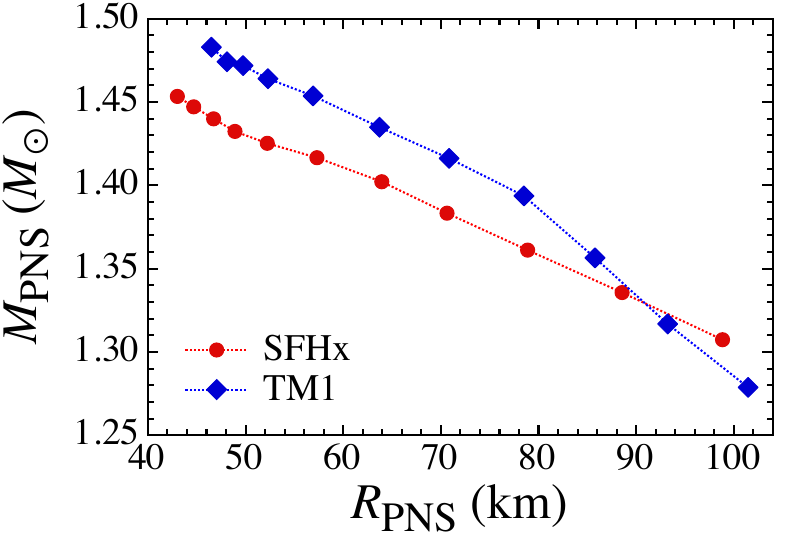}
 \caption{Evolution of protoneutron star mass and radius from 50 ms to 250 ms after core bounce, depending on EOS. The protoneutron star evolves from bottom-right to top-left in this figure. The simulations are conducted in 3D-GR. Taken from Ref.~\cite{SKTK2017}.}
\label{fig:MR_PNS}
\end{figure}

As seen in the top-left panel of Fig.~\ref{fig:rho_t_DD2}, the density is continuously distributed outward, and there is no obvious boundary expressing the stellar surface. Nevertheless, in a later phase, e.g., at 230 and 320 ms after core bounce, a sharp decrease in density is observed around $\rho\sim 10^{10}-10^{12}$ g/cm$^3$. So, in this study, we define the protoneutron star surface, where the density becomes the surface density given by $\rho_s\equiv 10^{11}$ g/cm$^3$. Thus, our protoneutron star models depend on the selection of the surface density, but, as shown in the next section, the oscillation frequencies of the protoneutron stars corresponding to the primary gravitational waves (ramp-up signal) in the numerical simulations hardly depend on the selection of the surface density. Namely, as far as we discuss the connection between the primary gravitational waves in the numerical simulation and the protoneutron-star oscillation frequencies, the uncertainty in the selection of the surface density is not important. Then, the mass ($M_{\rm PNS}$) and radius ($R_{\rm PNS}$) of the protoneutron stars are determined at each time step, while their evolution depends on the model parameters in the supernovae. For instance, in Fig.~\ref{fig:MR_PNS}, the relations of $M_{\rm PNS}$ and $R_{\rm PNS}$ with different EOSs, i.e., SFHx~\cite{SFHx} and TM1~\cite{TM1}, adopting the $15M_\odot$ progenitor star, are shown at each time step after core bounce, where the stellar models evolve from right-bottom to left-top direction~\cite{SKTK2017}. Therefore, if one can extract the temporal evolution of the mass and radius of a protoneutron star, it would provide crucial information for constraining the EOS.
We note the mass of the resultant neutron star depends on the progenitor mass, while the construction of neutron star models with almost the maximum mass from the numerical simulation for the core-collapse suernovae are quite difficult.

\section{Linear analysis}
\label{sec:LA}

For a given protoneutron star model, the oscillation frequencies are determined via linear analysis. In this article, we calculate the oscillation frequencies with two different approaches. One is a simple way, the so-called Cowling approximation, where the metric perturbations are neglected during the fluid oscillations. The perturbation equations can be derived by linearizing the relativistic energy-momentum conservation law. Since one considers only the fluid oscillations with this approach, the numerical domain to solve is only inside the star, where the imposed boundary conditions are the regularity condition at the center and the condition that the Lagrangian perturbation of pressure should be zero at the stellar surface. On the other hand, as a second approach, we also calculate the frequencies with metric perturbations, where the perturbation equations can be derived by linearizing the Einstein equations. With metric perturbations, one has to consider not only the region inside the star but also the exterior of the star. Thus, the boundary condition at the spatial infinity, where the gravitational waves should be only outgoing waves, is also imposed in addition to the conditions imposed on the Cowling approximation. The concrete perturbation equations and boundary conditions are shown in Ref.~\cite{SKTK2019} for the Cowling approximation and in Ref.~\cite{ST2020} for the case with metric perturbations. In both approaches, one can add the normalization condition at some point, since this is a linear analysis. Then, the problem to solve becomes an eigenvalue problem with respect to the eigen-angular-frequency, $\omega$. 
We note that the eigenfrequencies are determined by the value of azimuthal quantum number, $\ell$. However, in this manuscript we focus only on the quadrupole ($\ell=2$) modes, because those modes become dominant in the gravitational wave emission, e.g.,~\cite{Thone80}. 

\begin{figure}
 \centering
 \includegraphics[scale=0.5]{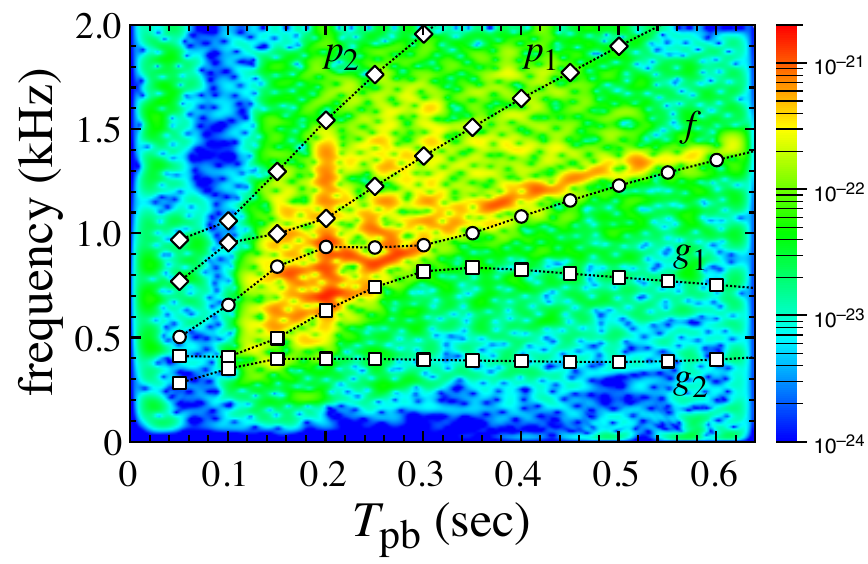}
 \caption{Comparison between the gravitational-wave signals appearing in the numerical simulation (contour) and oscillation frequencies of the protoneutron stars determined by solving the eigenvalue problem with the Cowling approximation (open marks). The simulation is conducted in 2D-effective GR with LS220 EOS. Taken from Ref.~\cite{ST2020a}.}
\label{fig:2D-spe}
\end{figure}

\begin{figure}
 \centering
 \includegraphics[scale=0.4]{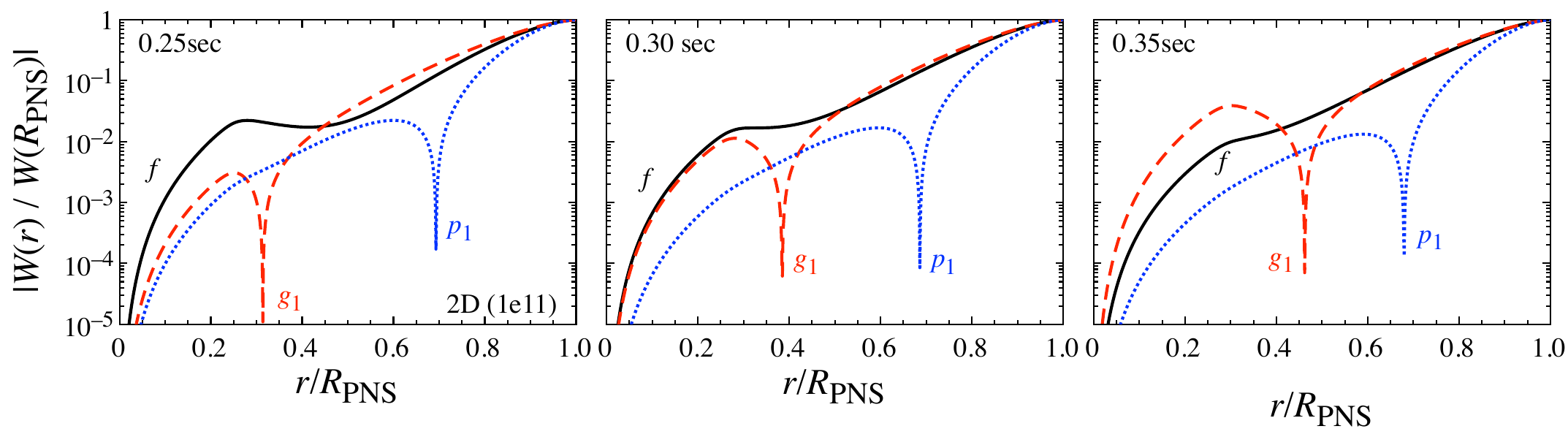}
 \caption{The radial profile of eigenfunctions of the $f$- (solid line), $p_1$- (dotted lines), and $g_1$-modes (dashed lines), normalized by the surface value. The left, middle, and right panels are the profiles at 0.25, 0.3, and 0.35 seconds after core bounce, shown in Fig.~\ref{fig:2D-spe}. Taken from Ref.~\cite{ST2020a}.}
\label{fig:Wr_AC}
\end{figure}

For instance, in Fig.~\ref{fig:2D-spe}, we show the correspondence between the gravitational-wave signals appearing in the effective GR simulation (background contour) and the specific frequencies of the protoneutron star at each time step determined by solving the eigenvalue problem with the Cowling approximation (open marks)~\cite{ST2020a},
where the progenitor model is the $2.9M_\odot$ helium star (He2.9) given in \cite{He2.9}, adopting LS220 EOS~\cite{LS220}. From this figure, one observes that the ramp-up signal in the simulation corresponds to the $g$-mode in the early phase and to the $f$-mode in the late phase, where the corresponding mode exchange via avoided crossing happens at $\sim 0.3$ seconds after core bounce. We note that the time when the avoided crossing emerges strongly depends on the model parameters. In this study, we classify the oscillation modes by counting the number of nodal points in the eigenfunction, following the standard mode classification. That is, the eigenfrequency without a node is the $f$-mode, while those lower (higher) than the $f$-mode frequencies are the $g_i$- ($p_i$-) modes, where the subscript $i$ denotes the nodal number.
We note that, although the correspondence between the eigenfrequencies and specific modes may depend on the mode classification, it has been shown that the mode spectrum becomes similar independently of the mode classification at least for the frequencies obtained with the Cowling approximation at later times ($\sim 0.4$ seconds) after the core-bounce~\cite{Rodriguez23}, based on the protoneutron star models obtained from 3D simulations of core-collapse supernovae conducted by Radice et al.~\cite{Radice19}.
In Fig.~\ref{fig:Wr_AC}, we show the radial profile of the eigenfunction, which is the radial displacement normalized by the value at the stellar surface, for the supernova model shown in Fig.~\ref{fig:2D-spe} at different times after core bounce, i.e., 0.25, 0.30, and 0.35 seconds, where the solid, dotted, and dashed lines denote the $f$-, $p_1$-, and $g_1$-modes, respectively. As seen in this figure, the $f$-mode has no node, while the $p_1$- and $g_1$-modes have one node.

\begin{figure}
 \centering
 \includegraphics[scale=0.5]{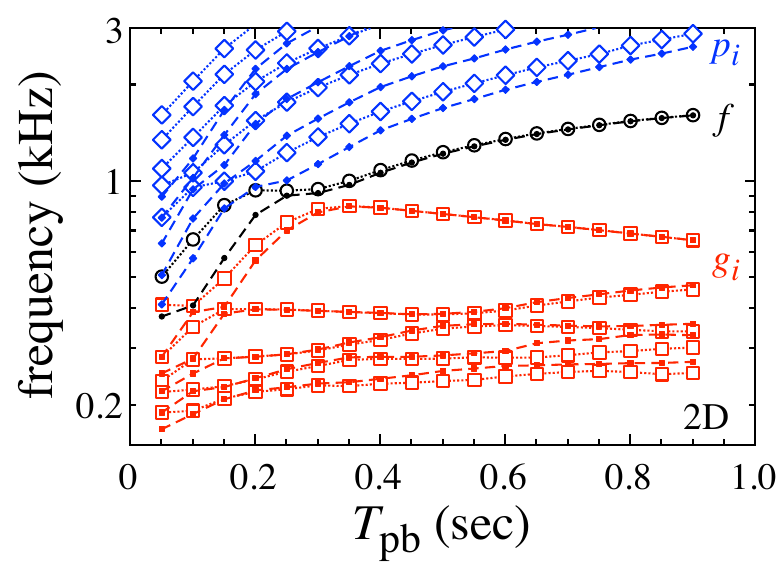}
 \caption{The eigenfrequencies excited in the protoneutron stars, defining the surface with the density $\rho_s=10^{11}$ g/cm$^3$ (dotted lines with open marks) and $10^{10}$ g/cm$^3$ (dashed lines with filled marks). The simulation is done in 2D-effective GR, using L220 EOS, while the linear analysis is done with the Cowling approximation. Taken from Ref.~\cite{ST2020a}.}
\label{fig:ft-2D}
\end{figure}

\begin{figure}
 \centering
 \includegraphics[scale=0.5]{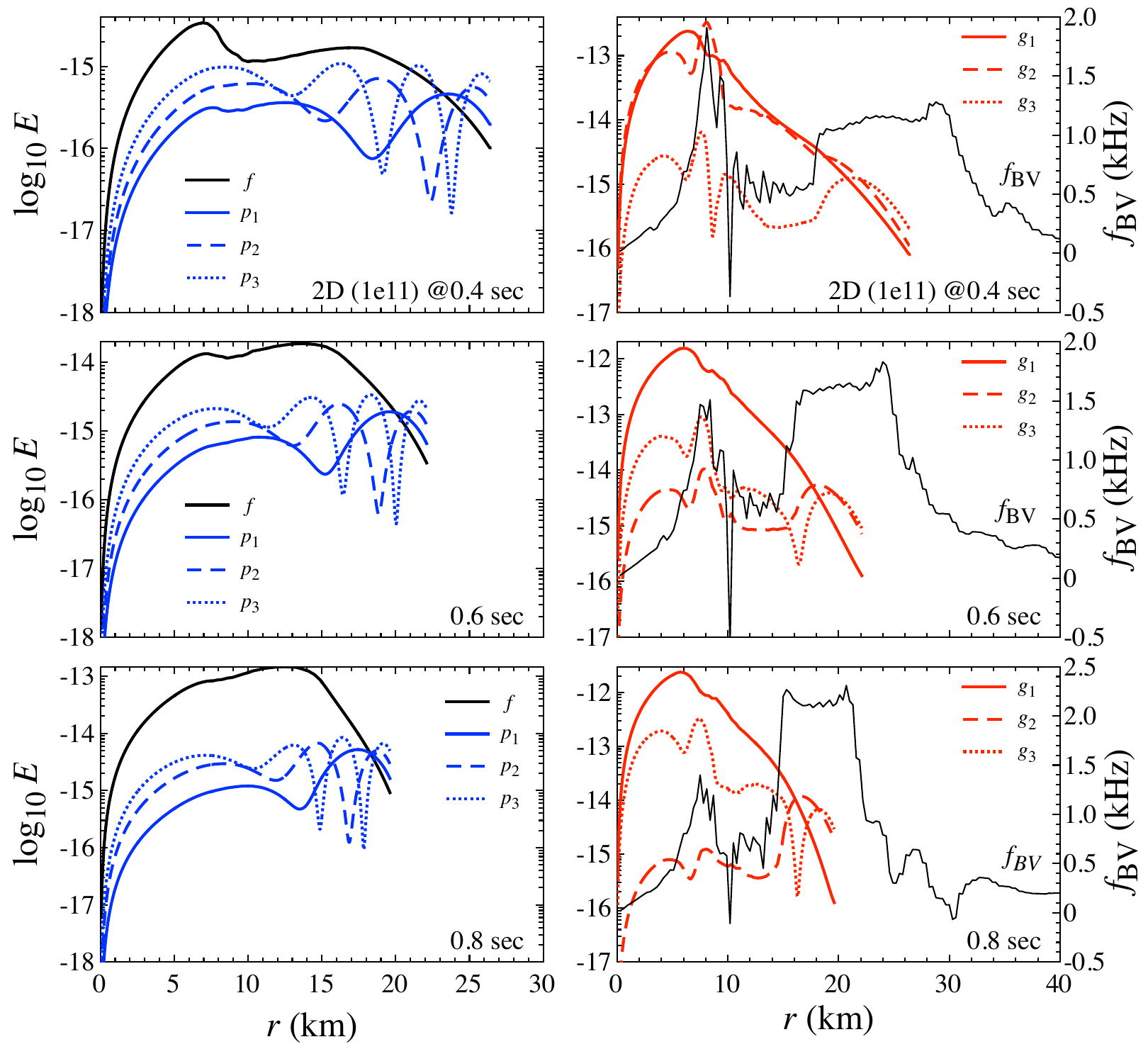}
 \caption{Radial profile of the pulsation energy, given by Eq.~(\ref{eq:pulsation_E}), for the $f$-, $p_i$-modes for $i=1,2,3$ in the left panel and for the $g_i$-mode for $i=1,2,3$ in the right panel. The radial profile of the Brunt-V\"{a}is\"{a}l\"{a} frequency, $f_{\rm BV}$, is also shown with a thin solid line in the right panel. The top, middle, and bottom panels correspond to 0.4, 0.6, and 0.8 sec. after core bounce for the supernova model shown in Fig.~\ref{fig:2D-spe}. Taken from Ref.~\cite{ST2020a}.}
\label{fig:Er_fg}
\end{figure}

As mentioned in the previous section, the protoneutron star models depend on the selection of the surface density, $\rho_s$. So, in principle, the oscillation frequencies also depend on the value of $\rho_s$. Nevertheless, as shown in Fig.~\ref{fig:ft-2D}, the $f$- and $g_1$-mode frequencies weakly depend on the selection of $\rho_s$, where the supernova model is the same as shown in Fig.~\ref{fig:2D-spe} \cite{MRBV2018,ST2020a}. 
This means that the selection of the surface density being $10^{11}$ g/cm$^3$ is reasonable for discussing the ramp-up signal in supernova gravitational waves.
This dependence may be understood, considering the pulsation energy density, $E(r)$, defined as 
\begin{equation}
  E(r) \simeq \frac{\omega^2 \varepsilon}{r^4}\left[W^2+\ell(\ell+1)V^2\right], \label{eq:pulsation_E}
\end{equation}
where $W$ and $V$ are the radial and angular displacements; $\omega$ is the eigenvalue (with which the eigenfrequency is given by $f=\omega / 2\pi$); and $\varepsilon$ is the background energy density. In Fig.~\ref{fig:Er_fg}, the radial profiles of pulsation energy for the supernova model shown in Fig.~\ref{fig:2D-spe} are plotted, where the left panel corresponds to the $f$- and $p_i$-modes for $i=1,2,3$ and the right panel is the $g_i$-modes for $i=1,2,3$. In the left panel, the radial profile of the Brunt-V\"{a}is\"{a}l\"{a} frequency, $f_{\rm BV}$, is also shown with the thin solid line. The top, middle, and bottom panels are different times after core bounce, i.e., 0.4, 0.6, and 0.8 seconds. From this figure, we find that the pulsation energy becomes dominant inside the star for the $f$- and $g_1$-mode oscillations, where the contribution from the region near the stellar surface is very small. Meanwhile, the pulsation energy for the other modes becomes dominant even near the stellar surface. The dependence of the eigenfrequencies on $\rho_s$ seems to be associated with this behavior of pulsation energy density. Moreover, from Fig.~\ref{fig:Er_fg}, we find that the radial position, $\sim 8$ km, where the $g$-mode becomes maxima, comes from the local maxima in the Brunt-V\"{a}is\"{a}l\"{a} frequency. In fact, the Brunt-V\"{a}is\"{a}l\"{a} frequency at this position decreases in time, which leads to the fact that the $g_1$-mode frequency decreases in time after avoided crossing with the $f$-mode. The pulsation energy density for the $g_2$- and $g_3$-modes additionally becomes large in the vicinity of the stellar surface, which is also associated with the distribution of the Brunt-V\"{a}is\"{a}l\"{a} frequency. 

\begin{figure}
 \centering
 \includegraphics[scale=0.5]{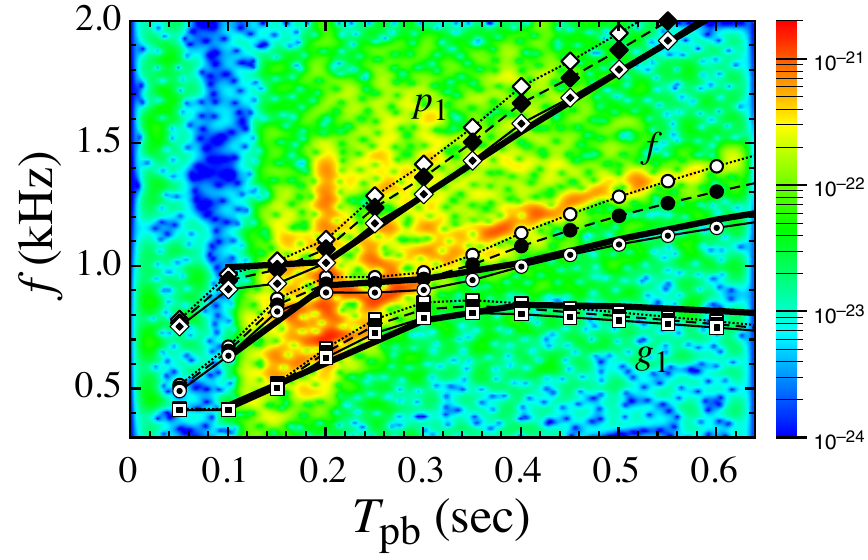}
 \caption{For the supernova model shown in Fig.~\ref{fig:2D-spe}, the frequencies of the protoneutron-star oscillations are shown with various approximations, and without approximation (with metric perturbations shown with thick solid lines). The open marks with dotted lines denote the results with the Cowling approximation; the filled marks with dashed lines are those including $\delta \alpha$ (allowing for perturbations to the lapse function $\alpha$); and the double marks with thin solid lines are those including $\delta \alpha$ and $\delta\psi$ (allowing for perturbations to $\alpha$ and the conformal factor $\psi$, but fixed the shift vector). The frequencies with $\delta \alpha$ and with ($\delta \alpha$ and $\delta \psi$) are calculate with GREAT~\cite{TCPOF19}. Taken from Ref.~\cite{ST2020}.}
\label{fig:2D-cow-full}
\end{figure}

Before discussing the universality of the supernova gravitational waves, we mention the accuracy of the Cowling approximation. For cold neutron stars, it is known that the Cowling approximation typically overestimates the frequencies at most $\sim 20\%$ \cite{YK97}. On the other hand, in the protoneutron star cases, we find that the behavior of the frequencies with the Cowling approximation is qualitatively the same as that without the approximation (with metric perturbations), and that the frequencies with the Cowling approximation can totally be determined within $\sim 20\%$ accuracy, where the $f$-mode with the Cowling approximation is overestimated~\cite{ST2020}. In Fig.~\ref{fig:2D-cow-full}, for the supernova model shown in Fig.~\ref{fig:2D-spe}, the frequencies of protoneutron stars determined with various approximations and also without approximation (with matrix perturbations) are shown. The open marks with dotted lines denote the frequencies determined with the Cowling approximation, i.e., the same as in Fig.~\ref{fig:2D-spe}; the filled marks with dashed lines are those partially taken into account the metric perturbation, i.e., the perturbation of the lapse function $\alpha$,  but the conformal factor $\psi$ and the shift vector $\beta^i$ are fixed; the double marks with thin solid lines are also those partially taking into account the metric perturbations, i.e., the perturbations of $\alpha$ and $\psi$, fixing $\beta^i$; and the thick solid lines denote the frequencies with the metric perturbations. The frequencies partially accounting for the metric perturbations are calculated with open code, GREAT~\cite{TCPOF19}. From this figure, if one compares the ramp-up signal in the simulation with the stellar frequencies determined with metric perturbations, their deviation becomes more significant, even though the frequencies with the metric perturbations are more realistic than those with the Cowling approximation. This point will be discussed in Sec.~\ref{sec:UR-GR}.

\section{Universal relations with effective GR simulations}
\label{sec:UR-eGR}

\begin{figure}
 \centering
 \includegraphics[scale=0.5]{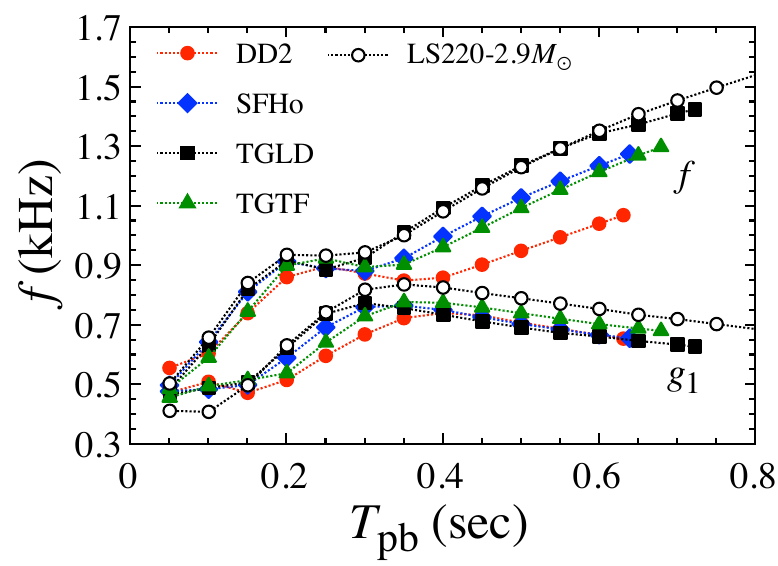}
 \caption{The evolution of protoneutron-star oscillation frequencies, focusing on the $f$- and $g_1$-modes, for various supernova parameters are shown as a function of the post-bounce time, $T_{\rm pb}$. The simulations are conducted in 2D-effective GR. Taken from Ref.~\cite{STT2021}.}
\label{fig:ft}
\end{figure}

After doing a similar analysis for the protoneutron-star oscillation frequencies calculated with the Cowling approximation, adopting various supernova models (in the effective GR), we confirm that the ramp-up signal in the simulation corresponds to the $g$-mode ($f$-mode) in the early (late) phase after core bounce in any model parameters we considered, although the systematic deviation between the ramp-up signal and protoneutron-star oscillation frequencies still emerges as in Fig.~\ref{fig:2D-spe}. If one focuses on the time evolution of the $g_1$- and $f$-mode frequencies, it depends on the supernova model parameters, as shown in Fig.~\ref{fig:ft}, where DD2~\cite{DD2}, SFHo~\cite{SFHx}, TGLD~\cite{TGLD}, and TGTF~\cite{TGTF} denote the models adopting the $20M_\odot$ progenitor model with different EOS, while LS220-2.9$M_\odot$ is the model adopting the $2.9 M_\odot$ helium star as a progenitor model with LS220 EOS~\cite{He2.9}. Furthermore, we note that the time evolution of the sequence of the $g_1$- to $f$-modes for the model with TGLD is different from that with TGTF, even though TGLD and TGTF are constructed with the same EOS properties for the uniform matter, where the treatment of nonuniform matter in the lower-density region is only different. Namely, the neutrino transport is sensitive to whether the nonuniform matter is constructed with the single nucleus approximation within the Thomas-Fermi model (TGTF) or the nuclear statistical equilibrium within the liquid-drop model (TGLD), which leads to a change in the gravitational-wave signals. 

\begin{figure}
 \centering
 \includegraphics[scale=0.5]{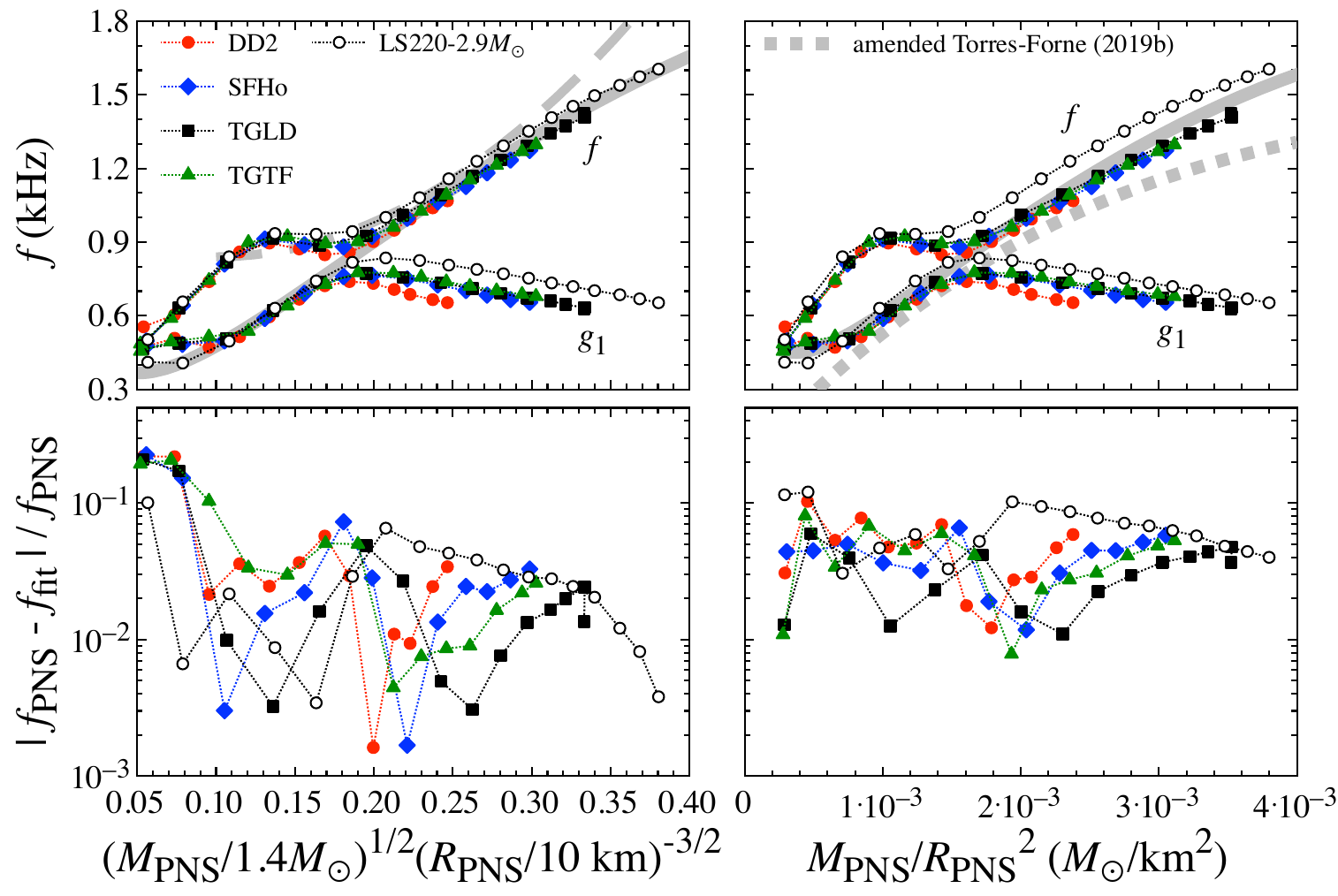}
 \caption{The top-left and top-right panels respectively correspond to the relations of the protoneutron-star oscillation frequencies as a function of the protoneutron star average density, $M_{\rm PNS}/R_{\rm PNS}^3$, and as a function of the protoneutron star surface gravity, $M_{\rm PNS}/R_{\rm PNS}^2$. The thick solid lines denote the universal relations given by Eq.~(\ref{eq:f_ave}) in the top-left panel and by Eq.~(\ref{eq:f_MR2}) in the top-right panel. We also show the universal relation given by Eq.~(\ref{eq:ff_BH}) for the $f$-mode oscillations in the case of the black hole formation proposed in~\cite{SS2019} with the thick dashed line in the top-left panel, and that given by Eq.~(\ref{eq:Torres-Forne}) proposed in~\cite{TCPOF19b} with the thick dotted line in the top-right panel.
 The top panel is taken from Ref.~\cite{STT2021}.
 In addition, the relative deviation of the $f$- and $g_1$-mode frequencies associated with the ramp-up signal in the supernova gravitational waves from the fitting formulae given by Eq.~(\ref{eq:f_ave}) and Eq.~(\ref{eq:f_MR2}) are respectively shown in the bottom-left and bottom-right panels. From this figure, one can observe that the empirical formula as a function of the surface density (as shown in the right panel) may relatively depend on the progenitor mass.}
\label{fig:fx-surf}
\end{figure}

Meanwhile, once the sequence of the $g_1$- to $f$-modes frequencies are plotted as a function of the protoneutron star properties, such as the average density given by $M_{\rm PNS}/R_{\rm PNS}^3$ or the surface gravity given by $M_{\rm PNS}/R_{\rm PNS}^2$, with the protoneutron star mass ($M_{\rm PNS}$) and radius ($R_{\rm PNS}$), one can see that model dependence becomes very weak, as shown in Fig.~\ref{fig:fx-surf}. In particular, the correlation between the sequence of the $g_1$- to $f$-modes (corresponding to the ramp-up signal in the simulations) and the square root of the average density is quite tight, as in the left panel of this figure. Using our data of the protoneutron-star oscillation frequencies, we derive the fitting formula for such a sequence, given as
\begin{eqnarray}
  f ({\rm kHz}) &=& -1.410 - 0.443 \ln(x) +  9.337x - 6.714x^2, \label{eq:f_ave} \\
  f ({\rm kHz}) &=& -0.0752 - 0.2600 \ln(\bar{x}_3) + 0.7446\bar{x}_3 - 0.0600\bar{x}_3^2, \label{eq:f_MR2}
\end{eqnarray}
where $x$ and $\bar{x}_3$ are the normalized average density of the protoneutron star defined as $x\equiv \left(M_{\rm PNS}/1.4M_\odot\right)^{1/2}\left(R_{\rm PNS}/10\ {\rm km}\right)^{-3/2}$ and the normalized surface gravity given by $\bar{x}_3\equiv \bar{x}/0.001$ with $\bar{x}\equiv M_{\rm PNS}/R_{\rm PNS}^2$ in the unit of $M_\odot$/km$^2$~\cite{STT2021}, i.e., the thick solid lines in the left and right panels correspond to Eqs.~(\ref{eq:f_ave}) and (\ref{eq:f_MR2}), respectively. In addition to our fitting formula derived here, for reference, the thick dashed line in the left panel is the fitting formula obtained in the case of black hole formations due to the massive progenitor~\cite{SS2019}, given by
\begin{equation}
  f_f ({\rm kHz}) = 0.9733 - 2.7171x + 13.7809x^2, \label{eq:ff_BH}
\end{equation}
while the thick dotted line in the right panel is the fitting formula proposed by the other group~\cite{TCPOF19b}, given by
\begin{equation}
  f ({\rm kHz}) = 620 \bar{x}_3 - 94.5 \bar{x}_3^2 + 5.30 \bar{x}_3^3. \label{eq:Torres-Forne}
\end{equation}
Unfortunately, the reason why our fitting formula (thick solid line) in the right panel deviates from that proposed in~\cite{TCPOF19b} (thick dotted line) is unclear, but at least Eq.~(\ref{eq:Torres-Forne}) deviates from the protoneutron-star oscillation frequencies obtained in our calculations. Maybe, this deviation comes from the fact that,  to derive the fitting formula in~\cite{TCPOF19b}, they considered the protoneutron star models provided with not only the 2D-simulations but also the 1D-simulations, and/or not only in the effective-GR but also in GR simulations. Anyway, as mentioned before, to express (or to discuss) the ramp-up signal in the supernova gravitational waves, the average density of the protoneutron star seems to be more suitable than the surface gravity.

Although we have shown that the correspondence between the ramp-up signal and protoneutron-star oscillation frequency in Fig.~\ref{fig:2D-spe}, we found a systematic deviation between the two frequencies by carefully observing them. This systematic deviation may come from the inconsistency of the gravitational theories for the numerical simulations and for the linear analysis, i.e., the numerical simulations have been conducted in the effective GR, while the linear analysis is done in the relativistic framework. To check this point, we will consider the protoneutron star asteroseismology, using the GR numerical simulation, in the next section.

\section{Universal relations with GR simulations}
\label{sec:UR-GR}

To understand the systematic deviation between the gravitational-wave signals in the effective-GR simulations and the protoneutron-star oscillation frequencies determined within the Cowling approximation, we now consider comparing the ramp-up signal with the protoneutron-star oscillation frequencies, using numerical data obtained from 2D-GR simulations with SFHo EOS~\cite{SFHx}. 

\subsection{2D-GR simulations with a monopole gravitational potential}
\label{sec:GR1}

\begin{figure}
 \centering
 \includegraphics[scale=0.5]{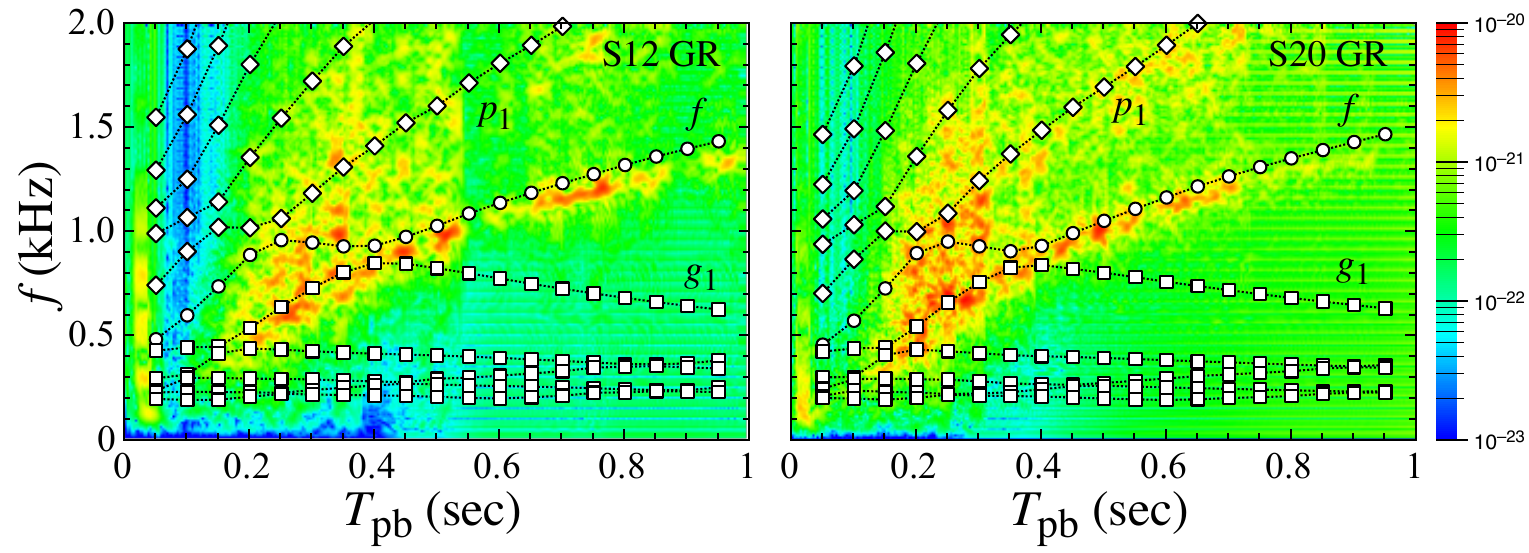}
 \caption{Comparison between the gravitational-wave signals in the 2D-GR simulation (contour), assuming the monopole gravitational potential, and the protoneutron-star oscillation frequencies with the Cowling approximation (open marks), where the left and right panels correspond to the results with the $12M_\odot$ and $20M_\odot$ progenitor models, adopting the SFHo EOS.  Taken from Ref.~\cite{{SMT24}}.}
\label{fig:GW_SDHO_GR}
\end{figure}

First, we consider the case with the 2D-GR simulations with a monopole gravitational potential. In Fig.~\ref{fig:GW_SDHO_GR}, we show the gravitational-wave signals in the 2D-GR simulations with monopole gravitational potential (contour) and the oscillation frequencies with the Cowling approximation on the protoneutron star models constructed from the same numerical simulations (open marks), where the left and right panels correspond to the results with the $12M_\odot$ and $20M_\odot$ progenitor models. Unlike the case of the effective-GR simulations as shown in Fig.~\ref{fig:2D-spe}, we find that the ramp-up signal in the GR-simulations seem to correspond well with the protoneutron-star oscillation frequencies. That is, the systematic deviation shown in Fig.~\ref{fig:2D-spe}, i.e., mismatch between the ramp-up signal in the effective-GR simulations and protoneutron-star oscillation frequencies, seems come from the inconsistency of the gravitational theories in the simulations and for the linear analysis. In addition, since the numerical simulations are conducted with a monopole gravitational potential, it is reasonable that the protoneutron-star oscillation frequencies calculated with the Cowling approximation agree well with the results of the numerical simulations. Meanwhile, the frequencies with the Cowling approximation can generally overestimate the frequencies, e.g., in Fig.~\ref{fig:2D-cow-full}. Thus, once one calculates the protoneutron-star oscillation frequencies with the metric perturbations (without the Cowling approximation), the stellar frequencies are expected to systematically deviate downward from the ramp-up signal in the simulations.  

\begin{figure}
 \centering
 \includegraphics[scale=0.5]{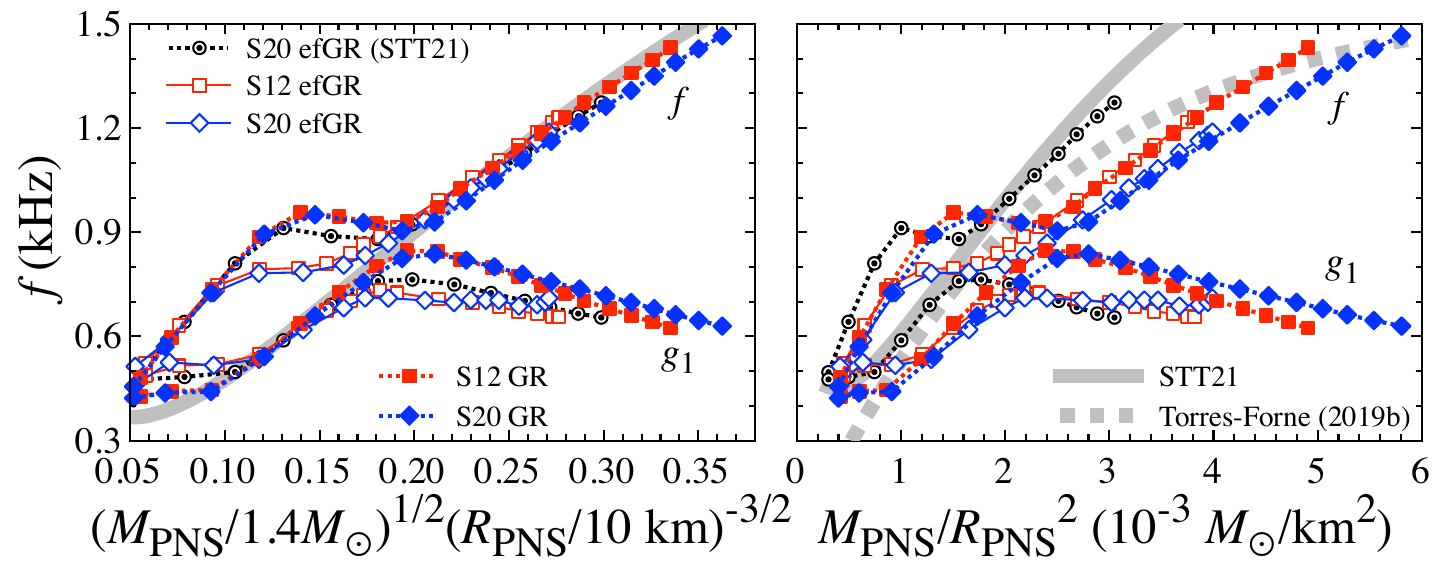}
 \caption{The $g_1$- and $f$-mode frequencies, corresponding to the ramp-up signal in the simulations, are shown as a function of the normalized protoneutron star average density in the left panel and the surface gravity in the right panel, adopting SFHo EOS. The thick solid lines in the left and right panels are the same as the fitting formula derived in Fig.~\ref{fig:fx-surf}, i.e., Eqs.~(\ref{eq:f_ave}) and (\ref{eq:f_MR2}), and the thick dotted line is also same as the fitting formula proposed in~\cite{TCPOF19b}, i.e., Eq.~(\ref{eq:Torres-Forne}). In the figure, the results denoted with STT21 (double circles) are the same as those with the filled diamonds shown in Fig.~\ref{fig:fx-surf}; the results with S12 efGR and S20 efGR (open marks) are those with the simulation in the effective GR, but the adopted numerical interpolation is different from that in the simulations discussed in the previous section; and the results with S12 GR and S20 GR (filled marks) are those with the GR 2D simulations shown in Fig.~\ref{fig:GW_SDHO_GR}. Taken from Ref.~\cite{SMT24}.}
\label{fig:Universal-GR}
\end{figure}

Focusing on the $g_1$- and $f$-mode frequencies, which correspond to the ramp-up signal in the simulations, in Fig.~\ref{fig:Universal-GR}, the oscillation frequencies on the protoneutron star models constructed using the 2D-GR simulations with a monopole gravitational potential shown in Fig.~\ref{fig:GW_SDHO_GR}, are shown as a function of the root square of the protoneutron star average density (left panel) and as a function of the surface gravity (right panel). In both panels, the data labeled S12 GR and S20 GR correspond to the results of the 2D-GR simulations with the $12M_\odot$ and $20M_\odot$ progenitor models. The data labeled S20 efGR are identical to those obtained using SFHo shown in Fig.~\ref{fig:fx-surf}, while those labeled S12 efGR and S20 efGR are also the results using the effective-GR simulations with SFHo EOS and with the $12M_\odot$ and $20M_\odot$ progenitor models, but the numerical interpolation adopted in the simulations is different from that adopted in the simulations discussed in the previous section (see~\cite{SMT24} for details). The thick solid lines are the fitting formulae given by Eqs.~(\ref{eq:f_ave}) and (\ref{eq:f_MR2}), and the thick dotted line is the fitting formula proposed in~\cite{TCPOF19b}, given by Eq.~(\ref{eq:Torres-Forne}). From this figure, we find that the protoneutron-star oscillation frequencies, which correspond to the ramp-up signal in the simulations, relatively depend on the gravitational theory in the simulations and also the numerical interpolation in the simulations, if one considers those as a function of the protoneutron-star surface gravity. On the other hand, we also find that they hardly depend on the gravitational theory and the numerical interpolation in the simulations, if one considers those as a function of the square root of the stellar average density, and confirm that, surprisingly, the oscillation frequencies with the Cowling approximation on the protoneutron star models constructed with the 2D-GR simulations can still be well expressed with a fitting formula derived from the results with the 2D-effective GR simulations, i.e., Eq.~(\ref{eq:f_ave}).

\subsection{2D-GR simulations with a non-monopole (2D) gravitational potential}
\label{sec:GR2}

\begin{figure}
 \centering
 \includegraphics[scale=0.5]{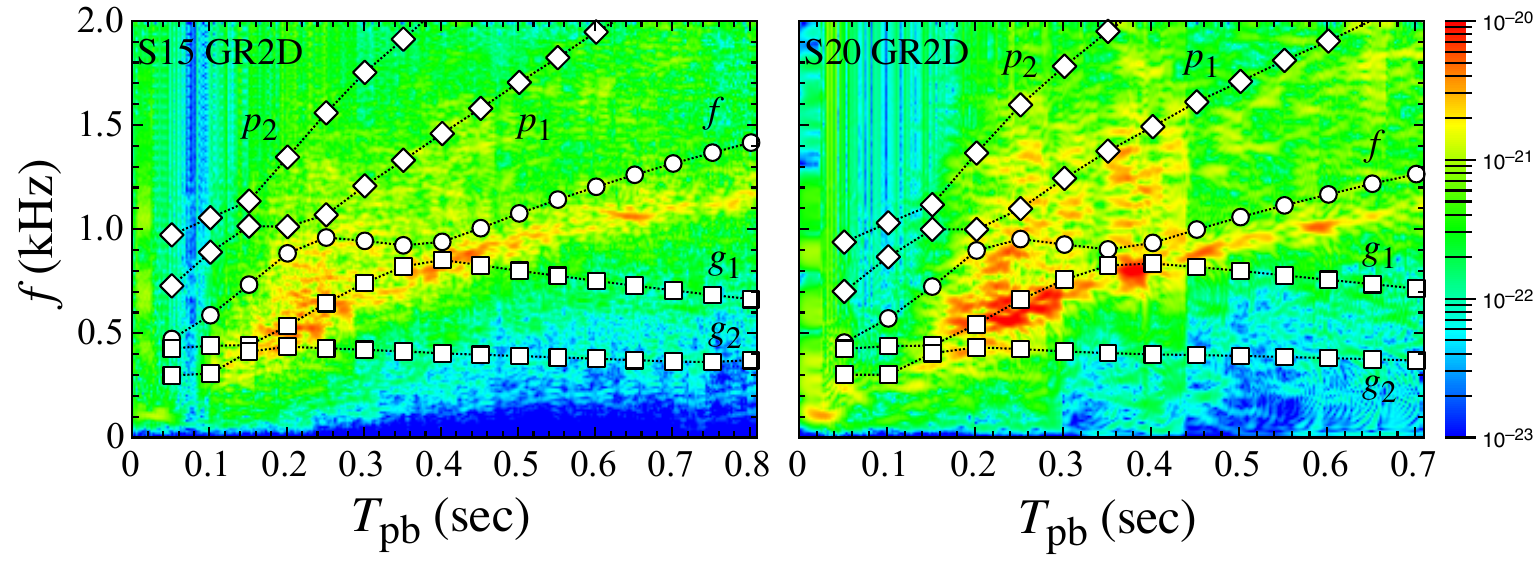}
 \caption{Same as Fig.~\ref{fig:GW_SDHO_GR}, i.e., the protoneutron-star oscillation frequencies are calculated with the Cowling approximation, but the simulations are conducted with non-monopole (2D) gravitational potential. The left and right panels correspond to the results for the $15M_\odot$ and $20M_\odot$ progenitors, respectively. Taken from Ref.~\cite{SMT25}.}
\label{fig:GW_SFHOB_2DM_cow}
\end{figure}

\begin{figure}
 \centering
 \includegraphics[scale=0.5]{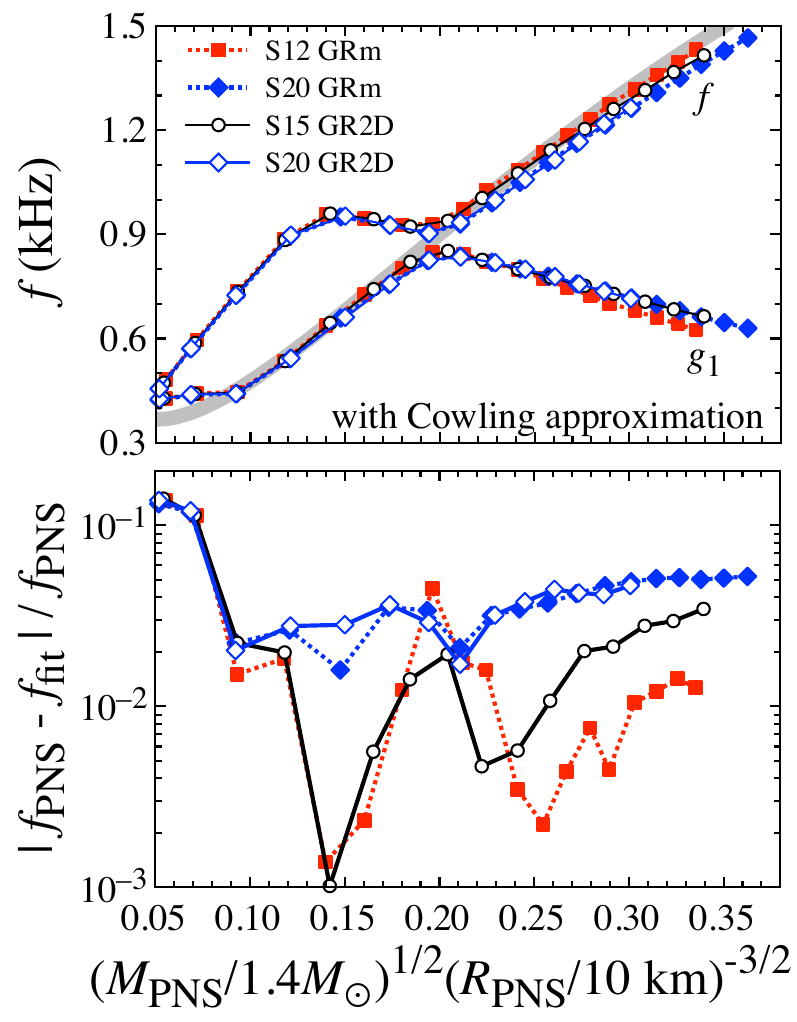}
 \caption{In the top panel, the $g_1$- and $f$-mode frequencies calculated with the Cowling approximation for the protoneutron star models constructed using the data of the 2D-GR simulations with the monopole gravitational potential (denoted by GRm) and the non-monopole (2D) gravitational potential (denoted by GR2D), are shown as a function of the square root of the normalized protoneutron star average density, where S12, S15, and S20 mean that the adopted progenitoar mass is $12M_\odot$, $15M_\odot$, and $20M_\odot$. That is, the results with S12 GRm and S20 GRm are the same as those with S12 GR and S20 GR in Fig.~\ref{fig:Universal-GR}. The thick solid line is the fitting formula derived from frequencies calculated with the Cowling approximation, using the data on effective GR situations, as given by Eq.~(\ref{eq:f_ave}). Top panel is taken from Ref.~\cite{SMT25}.
 In the bottom panel, we also show the relative deviation of the $f$- and $g_1$-mode frequencies excited in the protoneutron stars, associated with the ramp-up signals in the supernova gravitational waves, from the frequencies estimated with the fitting formula given by Eq.~(\ref{eq:f_ave}). Even though the fitting formula has been derived using the data for the effective GR, one can observe that it works well even for the case with the GR simulation with monopole gravity.  }
\label{fig:fx-universal}
\end{figure}

In the previous subsection, we considered the correspondence between the ramp-up signal in the 2D-GR simulations and the protoneutron-star oscillation frequencies with the Cowling approximation, adopting a monopole gravitational potential. However, in a more realistic situation, one should take into account the effect of a non-monopole gravitational potential in the simulations, and discuss the correspondence between the ramp-up signal in the simulations and the excited stellar oscillation frequencies. As a first step, we compare the gravitational wave signals in such simulations with the protoneutron-star oscillation frequencies determined with the Cowling approximation. The results are shown in Fig.~\ref{fig:GW_SFHOB_2DM_cow}, where the left and right panels correspond to the results with the $15M_\odot$ and $20M_\odot$ progenitor models, adopting the SFHo EOS~\cite{SMT25}. As a result of adopting a non-monopole gravitational potential, we find that the protoneutron-star oscillation frequencies become higher than the ramp-up signal in the simulations. Using the resultant stellar frequencies, we also check the universality as a function of the square root of the protoneutron star average density. Then, as shown in Fig.~\ref{fig:fx-universal}, we find that the oscillation frequencies on the protoneutron stars constructed with the 2D-GR simulations with a non-monopole gravitational potential, labeled with S15 GR2D and S20 GR2D, are still well expressed with the universal relation obtained from the results with the effective GR simulations, adopting the Cowling approximation, given by Eq.~(\ref{eq:f_ave}). We note that the results labeled with ``S12~GRm" and ``S20~GRm" in Fig.~\ref{fig:fx-universal} are respectively the same as those labeled with ``S12~GR" and ``S20~GR" in Fig.~\ref{fig:Universal-GR}, i.e., the eigenfrequencies excited in the protoneutron stars constructed with the 2D-GR simulations assuming a monopole gravitational potential.

\begin{figure}
 \centering
 \includegraphics[scale=0.5]{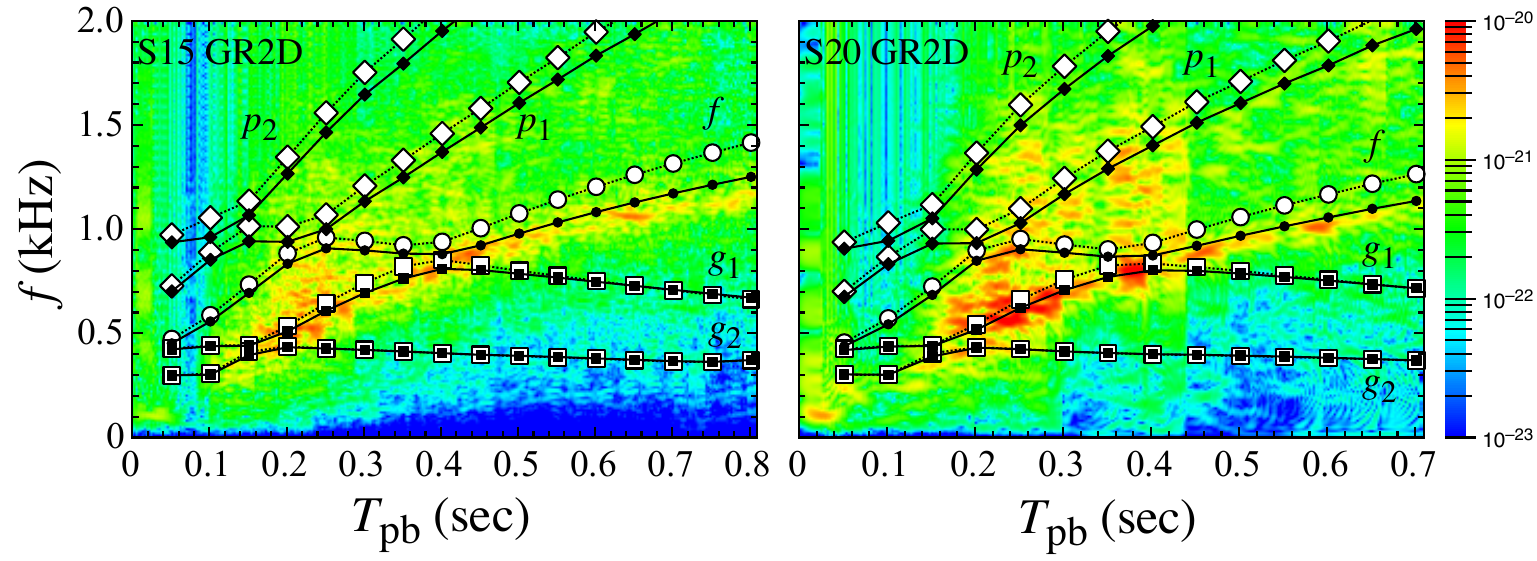}
 \caption{Same as in Fig.~\ref{fig:GW_SFHOB_2DM_cow}, i.e, the simulations are conducted in 2D-GR with the non-monopole (2D) gravitational potential, and the protoneutron-star oscillation frequencies are calculated with the Cowling approximation (open marks). In addition, the protoneutron-star oscillation frequencies calculated with the metric perturbations (without the Cowling approximation) are also shown with filled marks and solid lines. Taken from Ref.~\cite{SMT25}.}
\label{fig:GW_SFHOB_2DM}
\end{figure}

\begin{figure}
 \centering
 \includegraphics[scale=0.5]{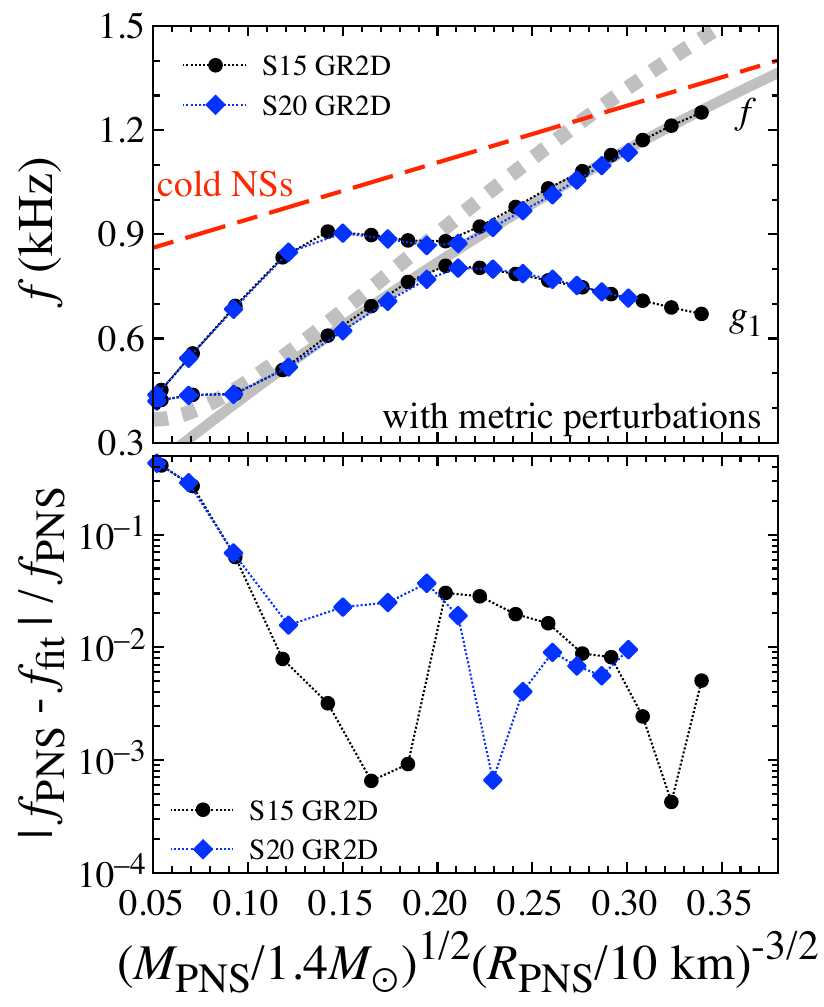}
 \caption{In the top panel, the $g_1$- and $f$-mode frequencies calculated with the metric perturbations are plotted as a function of the square root of the normalized protoneutron star average density, where the protoneutron star models are constructed using the data of the 2D-GR simulations with non-monopole (2D) gravitational potential. The thick solid line is a new fitting formula given by Eq.~(\ref{eq:univeral_new}), while the thick dotted line is the same as the fitting formula derived from the frequencies calculated with the Cowling approximation for the protoneutron star models constructed using the data of the 2D-effective GR simulations with monopole gravitational potential, given by Eq.~(\ref{eq:f_ave}). Furthermore, the thin-dot-dash-line denotes the frequencies estimated with the universal relation for the $f$-mode excited in the cold neutron stars given by Eq.~(\ref{eq:ff_cold}). The top panel is modified from the figure in Ref.~\cite{SMT25}. In the bottom panel, we also show the relative deviation of the $f$- and $g_1$-mode frequencies, $f_{\rm PNS}$, corresponding to the supernova gravitational waves from the frequencies estimated with Eq.~(\ref{eq:univeral_new}), $f_{\rm fit}$. One can see the fitting formula works well within a few \% accuracy for the protoneutron star models, whose square root of the average density is larger than $\sim 0.1$.}
\label{fig:fx-universal-new}
\end{figure}

Next, we consider comparing the ramp-up signal in the simulations with a non-monopole gravitational potential and the stellar oscillation frequencies determined with the metric perturbations. In Fig.~\ref{fig:GW_SFHOB_2DM}, we show the oscillation frequencies determined with the Cowling approximation (open marks), which are the same as shown in Fig.~\ref{fig:GW_SFHOB_2DM_cow}, and those with the metric perturbations (filled marks).  As expected, the frequencies with the metric perturbations become lower than those with the Cowling approximation, which corresponds to the fact that the Cowling approximation overestimates the frequencies. Anyway, we find that the stellar oscillations determined with the metric perturbations agree with the ramp-up signals in the simulation with a non-monopole gravitational potential. However, considering the situation that the oscillation frequencies with the Cowling approximation can be expressed well with the universal relation given by Eq.~(\ref{eq:f_ave}), as shown in Fig.~\ref{fig:fx-universal}, it is expected that the oscillation frequencies with the metric perturbations on the protoneutron star models constructed with the same numerical data obtained from the 2D-GR simulation with a non-monopole gravitational potential would deviate from this universality. In fact, as shown in Fig.~\ref{fig:fx-universal-new}, the frequencies of supernova gravitational waves hardly depend on the progenitor mass, and are still characterized by the protoneutron star average density, but those deviate from the universal relation derived from the results with the Cowling approximation, i.e., Eq.~(\ref{eq:f_ave}), shown with the thick dotted line. So, to express the supernova gravitational waves derived from the 2D-GR simulations with a non-monopole gravitational potential, we have to derive a new fitting formula:
\begin{equation}
   f_{\rm  2D}\ ({\rm kHz}) = 0.0082 + 4.5908x - 2.6821x^2, \label{eq:univeral_new}
\end{equation}
which are shown with the thick solid line in Fig.~\ref{fig:fx-universal-new}.
We note that the universal relation expressing the $f$-mode frequencies as a function of the average density has already been derived for cold neutron stars~\cite{AK96,AK98}, i.e., 
\begin{equation}
  f_f^{({\rm cold})}\ ({\rm kHz}) \simeq 0.78 + 1.635x \label{eq:ff_cold}.
\end{equation}
To compare this relation to our empirical relations, we also plot the frequencies estimated by Eq.~(\ref{eq:ff_cold}) in Fig.~\ref{fig:fx-universal-new} with a thin-dot-dash-line. From this comparison, one can find that the empirical relation for the protoneutron stars is obviously different from that for the cold neutron stars. This means that the empirical relation for the $f$-mode frequencies depends on the core-bounce time, which would converge to the relation for the cold neutron stars after the thermal evolution.

In Table~\ref{tab:fGW}, we summarize the correspondence between the ramp-up signals appearing in the 2D simulations, $f_{\rm GW}$, and the eigenfrequencies of the protoneutron stars corresponding to the ramp-up signals, $f_{\rm PNS}$, where the simulations are conducted with effective GR or GR, while the stellar oscillations are determined with the Cowling approximation or with the metric perturbations. In addition, the GR simulations are conducted with either a monopole or non-monopole (2D) gravitational potential. In fact, although we did not compare the ramp-up signals with the protoneutron-star oscillation frequencies with the metric perturbations, adopting the 2D-GR simulations with a monopole gravitational potential, we expect that $f_{\rm GW}$ would be higher than $f_{\rm PNS}$, considering the situation of $f_{\rm GW}\approx f_{\rm PNS}$ with the Cowling approximation and the Cowling approximation generally overestimates the frequencies compared to the frequencies with the metric perturbations. As we have already mentioned, the ramp-up signals in the GR simulations with a monopole (non-monopole) gravitational potential agree well with the protoneutron-star oscillation frequencies determined with the Cowling approximation (with the metric perturbations). 

\begin{table}
\caption{Comparison between the frequencies appearing in the 2D simulations, $f_{\rm GW}$, and eigenfrequencies of protoneutron stars determined with Cowling approximation or with metric perturbations, $f_{\rm PNS}$, for different treatments of gravity in the simulations. The number in the bracket denotes the dimension of gravity in the simulations. }
\centering
\begin{tabular}{ccc}
\hline
\hline
gravity in simulations & Cowling approximation & with metric perturbations  \\
\hline
Effective GR (1)  & $f_{\rm GW}>f_{\rm PNS}$ &  $f_{\rm GW}>f_{\rm PNS}$ \\
GR (1)  & $f_{\rm GW}\approx f_{\rm PNS}$ & ($f_{\rm GW}>f_{\rm PNS}$) \\
GR (2)  & $f_{\rm GW}< f_{\rm PNS}$ & $f_{\rm GW}\approx f_{\rm PNS}$ \\
\hline
\hline
\end{tabular}
\label{tab:fGW}
\end{table}

\begin{figure}
 \centering
 \includegraphics[scale=0.5]{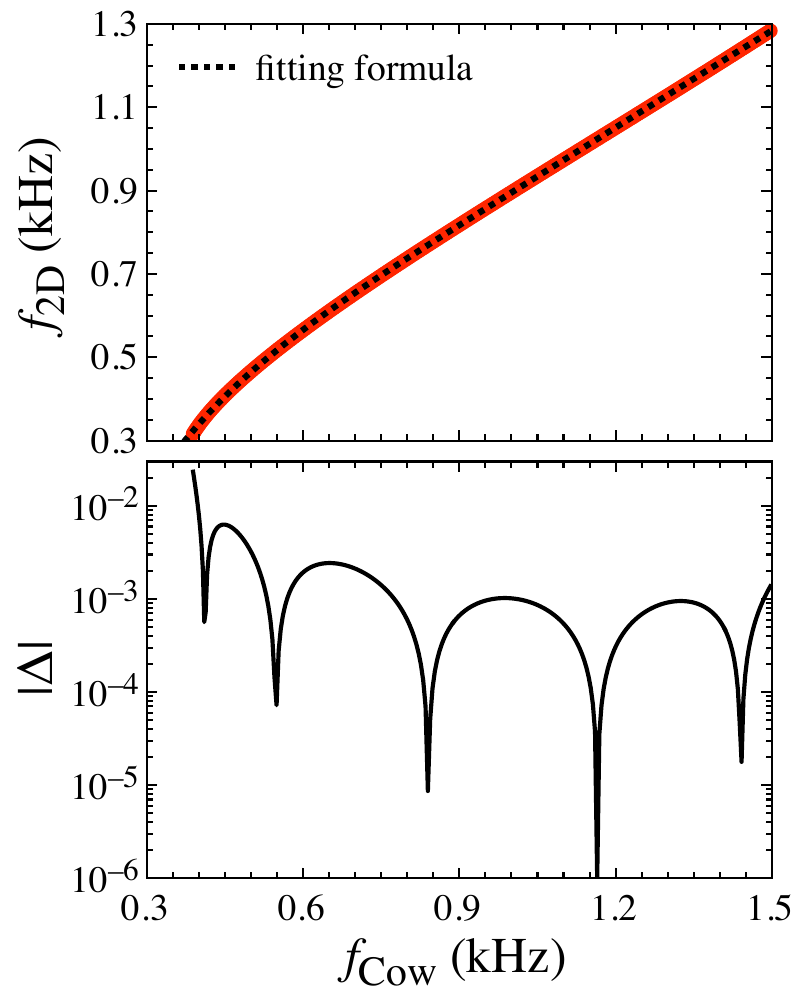}
 \caption{The oscillation frequencies calculated with the metric perturbations, $f_{\rm 2D}$, for the protoneutron star models constructed with the 2D-GR simulations with non-monopole (2D) gravitational potential are shown as a function of the frequencies with the Cowling approximation, $f_{\rm Cow}$, for the protoneutron star models constructed with the 2D-GR simulations with monopole potential. The dotted line is the fitting line given by Eq.~(\ref{eq:f2D-fCow}). In the bottom panel, the relative deviation of $f_{\rm 2D}$ from the fitting formula, calculated with Eq.~(\ref{eq:Delta}). Taken from Ref.~\cite{SMT25}.}
\label{fig:f2D_fCow_Delta}
\end{figure}

Finally, we propose a fitting formula to evaluate the gravitational-wave frequencies of the rump-up signal in GR-2D simulations with a non-monopole gravitational potential, using those determined with a monopole potential. As mentioned before, the frequencies of the rump-up signal obtained from the GR-2D simulations with a non-monopole (monopole) gravitational potential agree with the frequencies determined with the metric perturbations, $f_{\rm 2D}$, (with the Cowling approximation, $f_{\rm Cow}$) of the protoneutron stars constructed with the corresponding simulations, while $f_{\rm 2D}$ ($f_{\rm Cow}$) are well expressed as a function of the square root of the protoneutron-star average density, independently of the supernova model parameters, given by Eq.~(\ref{eq:univeral_new}) (Eq.~(\ref{eq:f_ave})). Using two fitting formulae, Eqs.~(\ref{eq:univeral_new}) and~(\ref{eq:f_ave}), we can derive the fitting formula to evaluate $f_{\rm 2D}$ using $f_{\rm Cow}$~\cite{SMT25}, such as
\begin{equation}
   f_{\rm  2D}^{\rm fit}\ ({\rm kHz}) = 1.7800 + 0.9676 \ln(f_{\rm Cow}) - 1.8052 f_{\rm Cow} + 1.1441 f_{\rm Cow}^2 - 0.2236 f_{\rm Cow}^3, \label{eq:f2D-fCow}
\end{equation}
where the unit of $f_{\rm Cow}$ is in kHz. In the top panel of Fig.~\ref{fig:f2D_fCow_Delta}, we show the values of $f_{\rm 2D}$ and those estimated with Eq.~(\ref{eq:f2D-fCow}) are shown as a function of $f_{\rm Cow}$, for the protoneutron star models with $x=0.07-0.35$, and the bottom panel is the relative deviation estimated with
\begin{equation}
   \Delta \equiv \frac{|f_{\rm  2D}^{\rm fit} - f_{\rm  2D}|}{f_{\rm  2D}}. \label{eq:Delta}
\end{equation}
From this figure, the fitting formula given by Eq.~(\ref{eq:f2D-fCow}) works well, with which one can estimate $f_{\rm 2D}$ from the results obtained by the 2D-GR simulations with a monopole gravity.

\section{Conclusion}
\label{sec:conclusion}

Core-collapse supernovae are one of the most promising sources of gravitational waves following the merger of compact binaries. However, gravitational waves from supernovae are far weaker than those from compact binary mergers, and gravitational waves detectable by current detectors are likely limited to those originating from supernovae within the Milky Way. So, we have to thoroughly prepare to extract physical properties for when a supernova explosion actually occurs. The most important signal among the supernova gravitational waves is the ramp-up signal, which appears in most of the simulations. The gravitational wave frequencies increase from hundreds of hertz to the kilohertz range within around one second after core bounce. The time evolution of the ramp-up signal definitely depends on the supernova parameters, such as the progenitor mass, EOS for dense matter, and the numerical methods, although the dependence of the numerical method is not physical. To extract the physical properties from the gravitational wave signals, avoiding such uncertainties, the universal relation(s), if any, are very crucial, which is the relation between the gravitational wave frequencies and physical properties independently of the supernova parameters. 

In this article, we particularly focused on the oscillation frequencies of protoneutron stars constructed using the numerical simulations, and compared them with the frequencies of ramp-up signal appearing in the 2D simulations, where the simulations are conducted with either the effective GR or GR, and the protoneutron-star oscillation frequencies are determined either with the Cowling approximation or with the metric perturbations. These simulations are conducted with a monopole gravitational potential, but in the case of the 2D-GR simulations, we also consider the simulations with a non-monopole (2D) gravitational potential. Then, we find that the ramp-up signal in the simulations with the effective GR, GR with a monopole potential, and GR with a non-monopole potential becomes higher, comparable, and lower than the protoneutron star frequencies with the Cowling approximation, while the ramp-up signal in the GR simulation with a non-monopole gravitational potential is comparable to the protoneutron star frequencies with metric perturbations. Even though how well the ramp-up signal in the simulations agrees with the protoneutron star frequencies depends on the gravitational theory and the dimension of the potential, the protoneutron star frequencies corresponding to the ramp-up signal, i.e., the $g$-modes in the early phase and the $f$-mode in the later phase after core bounce, determined with the Cowling approximation are well expressed as a function of the square root of the protoneutron star average density, independently of the gravitational theory and the dimension of the potential, which is given by Eq.~(\ref{eq:f_ave}). On the other hand, the protoneutron star frequencies with the metric perturbations deviate from this fitting formula, but they are still characterized by the square root of the protoneutron star average density, given by Eq.~(\ref{eq:univeral_new}). Furthermore, we also derive a fitting formula to estimate the ramp-up signals in the 2D-GR simulations with a non-monopole potential, using the results with a monopole potential, given by Eq.~(\ref{eq:f2D-fCow}). Since our study may not be enough to see the universality, especially for the case with a non-monopole gravity, we should test the validity of our universal relation, considering more supernova parameters.



\section*{Acknowledgments}
This work is supported in part by Japan Society for the Promotion of Science (JSPS) KAKENHI Grant Numbers 
JP23K20848         
and JP24KF0090.







\end{document}